\documentstyle[proceedings]{kluwer}
\begin{opening}
\title{RESULTS FROM THE OPTICAL GRAVITATIONAL LENSING EXPERIMENT (OGLE)}
\author{B. Paczy\'nski}
\author{K. Z. Stanek}
\institute{Princeton University Observatory\\
Princeton, NJ 08544--1001}
\author{A. Udalski}
\author{M. Szyma\'nski}
\author{J. Ka\L u\.zny}
\author{M. Kubiak}
\institute{Warsaw University Observatory \\
Al. Ujazdowskie 4, 00--478 Warszawa, Poland}
\author{M. Mateo}
\institute{Department of Astronomy, University of Michigan \\
821 Dennison  Bldg., Ann Arbor, MI~48109--1090}
\author{W. Krzemi\'nski}
\institute{Carnegie Observatories, Las Campanas Observatory \\
Casilla 601, La Serena, Chile}
\author{G. W. Preston}
\institute{Carnegie Observatories, Institution of Washington \\
813 Santa Barbara Street, Pasadena, CA 91101}
\end{opening}
\runningtitle{RESULTS FROM THE OGLE COLLABORATION}

\begin{document}

\begin{abstract}
The analysis of the first three years of the OGLE data revealed 12
{microlensing} events of the {Galactic bulge} stars, with the
characteristic time scales in the range $ 8.6 < t_0 < 80 $ days, where
$ t_0 = R_E / V $. A complete sample of nine events gave the
optical depth to {gravitational microlensing}
larger than $(3.3 \pm 1.2) \times 10^{-6}$, in excess of current
theoretical estimates, indicating a much higher efficiency for
microlensing by either bulge or disk lenses. The
lenses are likely to be ordinary stars in the {Galactic bar}, which has
its long axis elongated towards us. At this time we have no
evidence that the OGLE events are related to {dark matter}.
The OGLE {color magnitude diagrams} reveal the presence of the Galactic
bar and a low density inner disk region $ \sim 4 $ kpc in radius.
A catalogue of a few thousand {variable stars} is in preparation.
\end{abstract}

\section{INTRODUCTION}

The OGLE project (Optical Gravitational Lensing Experiment)
is a collaboration between the Warsaw University
Observatory, Carnegie Observatory, and Princeton University Observatory.
All observations are done with the 1 meter Swope telescope at the Las
Campanas Observatory, operated by the Carnegie Institution of Washington.
We have the telescope available to the OGLE about 70 nights each season,
which lasts from early April to early September, when the
{Galactic bulge}
can be observed.  The detector is a single Loral CCD with $(2k)^2$ pixels.
A somewhat modified DoPhot photometric software (Schechter, Mateo \&
Saha 1993) is used to extract stellar magnitudes from the CCD frames.
All data is processed at the observing site with a Sparc 10/512 computer
within 24 hours of the acquisition, and the summary of the results is
e-mailed to the Warsaw University Observatory for evaluation, off-loading
the observer who is on duty.
All technical details and the journals of the observations
are provided in Udalski et al.~(1992, 1994a).
The project covered three observing seasons: 1992, 1993, and 1994,
and it is likely to continue in the same form through 1995.

Some time in 1995 a new 1.3 meter R/C telescope will be
put into operation at the Las Campanas site.
It will be managed by the Warsaw University
Observatory, and it will be dedicated to massive {CCD photometry},
the search for gravitational microlensing being one of the prime
projects. Two other large collaborations are involved in a search
for gravitational microlensing of the LMC stars as well as the Galactic
bulge stars (Alcock et al.~1993; Aubourg et al.~1993; see also these
Proceedings).

\begin{figure}[t]
\vspace{8.5cm}
\includegraphics{ogle_fig1.ps}
\caption{The locations of the 11 OGLE lensing events (large circles)
are shown in the color -- magnitude diagram for Baade's Window
and the Galactic bar fields combined. Only $ \sim 5\% $ of all
stars are shown. The dashed line shows the limit of lenses detectability given
by the condition $I<19.5$. A calibrated photometry of one of the 12 OGLE
events is not available at this time.}
\end{figure}

\section{RESULTS}

Whenever Galactic bulge was observable and the seeing was adequate
we were monitoring 13 fields in the bulge region in 1992, and 20 fields in
1993 and 1994, each field being $ \sim 15' \times 15' $ in size.
The distribution of the fields in the sky is shown in
Stanek et al.~(1994b, cf. Fig.1., these Proceedings)
The total of 12 lensing events was discovered: 7 in 1992,
3 in 1993 and 2 in 1994 (Udalski et al. 1993a, 1994b,c,d,e).
The location of 11 of those lenses in the color magnitude diagram (CMD)
is shown in Fig.1.  It is clear that the lenses are scattered
over the part of the diagram where the Galactic bulge stars are
found.

A careful statistical study of a complete sample of nine
events found in the data from 13 fields observed in 1992 and 1993
confirmed the impression that the distribution of those
nine events was random in the CMD, and their amplitude distribution
was random as well, allowing for known biases in the detection
procedure (Udalski et al. 1994e).  In the same paper
the optical depth to microlensing was found to be
larger than $ ( 3.3 \pm 1.2 ) \times 10^{-6}$, in excess of original
theoretical estimates, indicating a much higher efficiency for
microlensing by either bulge or disk lenses.  We argued that the
lenses are likely to be ordinary stars in the Galactic bar, which has
its long axis somewhat inclined to the line of sight (Paczy\'nski et al.
1994b). Two examples of the microlensing events are presented in Fig.2.

\begin{figure}[p]
\vspace{9.3cm}
\includegraphics{ogle_fig2a.ps}
\vspace{9.5cm}
\includegraphics{ogle_fig2b.ps}
\caption{Examples of two microlensing events: long-lasting (upper panel)
and short-lasting (lower panel) (Udalski et al.~1994b,e).}
\end{figure}
\begin{figure}[t]
\vspace{8cm}
\includegraphics{ogle_fig3.ps}
\caption{The second microlensing event detected by the OGLE Early Warning
System and observed in real time (Udalski et al.~1994c).  The dashed
vertical line marks the date the event crossed the detection threshold --
August 20, 1994.  The first event ever to be detected in real time,
OGLE \#11, crossed the detection threshold on July 8, 1994.}
\end{figure}
\begin{figure}[t]
\vspace{10cm}
\includegraphics{ogle_fig4.ps}
\caption{OGLE \#7 -- the first clear case of lensing by a
binary system (Udalski et al.~1994d).}
\end{figure}

Another major highlight was the development and implementation in the 1994
season of the ``early warning system'', EWS (Paczy\'nski 1994a, Udalski
et al.  1994c).  The data from every
clear night were automatically analyzed by the on-site computer system
(Sparcstation 10/512), and the summary of results was automatically
forwarded by e-mail to Warsaw.  Two lensing events, OGLE \#11 and
OGLE \#12 (Fig.3), were detected on their rise on July 8 and
August 20, 1994, respectively, making it possible to follow
them up in detail.

The first fairly spectacular case of a microlensing by a double star,
OGLE \#7, was found in the 1993 data (Udalski et al. 1994d).  The observed
light curve, shown if Fig.4,
exhibits a sharp peak due to the source star crossing an optical
caustic in the lensing system.  Theoretical model predicted that a second
sharp peak should be present in the gap of the OGLE data.  Following the
OGLE discovery of the event \#7 the MACHO collaboration found the
object in their data, and the presence of the second sharp peak of
the light curve has been confirmed (Ch. Alcock, private communication).
Thanks to the ``early warning system'' it will be possible in the future
to obtain very good time coverage of such peaks in the light curves of
double lensing events, which in turn will allow detailed studies of normally
very faint stars.  Events like that resolve the disks of the lensed stars
just like stellar eclipses do.  Brightening by more than five magnitudes is
possible in some cases.

The {color magnitude diagrams} of the bulge region (Udalski et al. 1993b)
turned out to be very
useful source of information about the galactic structure.  Stanek et al.
(1994a,b) found that the red clump stars (the cloud of points around
$V-I=1.9, V=17$ in Fig.1) in the Galactic bulge were $ 0.37 \pm 0.025 $
mag brighter at (l,b) = $(+5.5^o,-3.5^o)$ than at $(-5.0^o,-3.5^o)$,
indicating the bulge is shaped as a bar, with the end at positive
galactic longitude being closer to us.

The distribution of the {galactic disk stars} turned out to be
unexpected.
The disk main sequence stars form a surprisingly narrow band, traceable
from $(V-I,V)=(0.9, 15)$ to $(1.4, 19)$ in Fig.1.  The analysis of the
observed distribution pointed to a rapid drop in the number density of
disk stars at a distance $ \sim 3 $ kpc from the Sun (Paczy\'nski et al.
1994a).  These are the disk stars which may contribute to the lensing of
the bulge stars, and the structure is apparent down to at least
$ M_V \approx 7 $, i.e. it includes the old disk population.  The
evidence that the inner disk may have very few moderate age stars
was available for some time (Baud et al. 1981, Blommaert et al. 1994).

In a systematic search for {periodic variable stars} thousands were
found,
and the catalogues are in preparation using the database of the OGLE
photometry (Szyma\'nski \& Udalski 1993).  These are mostly contact binaries,
other eclipsing binaries, {RR Lyrae} and SX Phenicis stars.

When the Galactic bulge could not be observed we had a number of secondary
targets.  In 1992 a few dozen clusters of galaxies were monitored for
about four months in an attempt to detect supernovae -- none were found.

In 1993 one field in {Sculptor} {dwarf galaxy} was monitored
and over 200
RR Lyrae stars were detected (Ka\l u\.zny et al 1994). Some examples
are shown in Fig.5. A correlation
was detected between the period and average V magnitude of RRab type
variables.  Also, two globular clusters, {47 Tuc} and
{Omega Cen}, were
monitored in a search for detached eclipsing binaries.  At least one
such system was found in Omega Cen, approximately one magnitude above
the main sequence turn-off point (Ka\l u\.zny et al., in preparation).

\begin{figure}[p]
\vspace{18cm}
\includegraphics{ogle_fig5.ps}
\caption{Examples of variable stars found in Sculptor dwarf galaxy
(Ka\l u\.zny et al.~1994) in $V\;mag$-phase diagrams.}
\end{figure}

In 1994 Omega Cen and 47 Tuc were monitored, and in addition some frames
of one field in the recently discovered {Sagittarius} dwarf (Ibata,
Gilmore and Irwin 1994) were taken.  We detected seven RR Lyrae variables
belonging to the dwarf, and these were used to determine the distance:
$ 25.2 \pm 2.8 $ kpc (Mateo et al. 1994).  The color magnitude diagram
was used to estimate the age to be $ \sim 10 $ Gyr.

Preliminary results from the OGLE were presented at a number of conferences
(Szyma\'nski et al. 1993, Udalski et al. 1992b, 1993c)

\section{OGLE RELATED PAPERS}

There was a number of papers written as a consequence of the OGLE observations.
Kiraga \& Paczy\'nski~(1994) noticed that the major contribution to the
optical depth to {gravitational microlensing} of the Galactic bulge
stars is from the Galactic bulge stars themselves. Kiraga~(1994) presented
theoretical maps of the optical depth as a function of galactic coordinates.
The high optical depth found by the OGLE was reproduced theoretically by Zhao,
Spergel and Rich (1994) with their dynamical model of the Galactic bar,
strengthening the case for the presence of the bar and for the bar origin of
the majority of OGLE lensing events.  Stanek (1994) noticed that the `red
clump' stars that are lensed should be systematically fainter than the general
population of the {red clump stars}, and the shift in the median
magnitude is directly proportional to the radial depth of the bulge/bar.

In general there is no unique relationship between the observed time scale
of a microlensing event $ t_0 $ and the lens mass (e.g. Kiraga and
Paczy\'nski 1994).  However, there is a case where the relation between
$ t_0 $ and the lens mass is simple and unique.  This is lensing of
the Galactic bulge stars and LMC stars by {globular clusters} which
are in front of them (Paczy\'nski 1994b).  This may be the most robust
(though rather time consuming) way to determine the mass function
of stars and brown dwarfs of which the globular clusters are made.

Renzini (1994) noticed that the very high number of the red clump stars
observed by the OGLE in the Galactic bulge can be explained with the
enhanced helium abundance of those stars, the enhancement related
to their high heavy element content.

Mao and Di Stefano (1995) developed a code to fit theoretical double
lens light curves to the data, and noticed that the OGLE \#6 event
might be due to a double lens.  The code was used to fit the
observations of the OGLE \#7 (Udalski et al. 1994d).

Many theoretical interpretation papers about the OGLE and
MACHO events were written, but they are beyond the scope of this
mini-review.

\section{OGLE BY INTERNET}

Since the beginning of the 1994 observing
season the OGLE project is capable of near real time
data processing (Paczy\'nski 1994, Udalski et al. 1994c).
The new computer system automatically
signals the events while they are on the rise, making it possible to carry out
photometric and/or spectroscopic follow-up observations. The observers who
would like to be notified about the on-going events should send their request
to A.~Udalski (udalski@sirius.astrouw.edu.pl).  Two microlensing events have
already been discovered with this system (Udalski et al. 1994c).

The photometry of the OGLE microlensing events, their finding charts,
as well as a regularly updated OGLE status report, including more information
about the ``early warning system'', can be found over Internet from
``sirius.astrouw.edu.pl'' host (148.81.8.1), using the ``anonymous ftp''
service (directory ``ogle'', files ``README'', ``ogle.status'',
``early.warning''). The file ``ogle.status''
contains the latest news and references to all OGLE related
papers, and the PostScript files of some publications, including
Udalski et al.~(1994c,e). The OGLE results are also available over
``World Wide Web'': ``http://www.astrouw.edu.pl/''.

This project was supported with the NSF grants AST
9216494 and AST 9216830 and Polish KBN grants No 2-1173-9101 and BST438A/93.

\end{document}